\documentclass[review]{elsarticle}

\usepackage{mathtools}
\usepackage{amsmath}
\usepackage{graphicx}
\usepackage{subcaption}
\usepackage{float}
\usepackage{tabularx}

\usepackage{lineno,hyperref}
\modulolinenumbers[5]

\begin{document}
	\nolinenumbers
	\let\linenumbers\nolinenumbers\nolinenumbers % just for arvix
	\begin{frontmatter}
		
		\title{Exponential and algebraic decaying solitary waves and their connection to hydraulic fall solutions}
		
		%\author{Keith C.H. Chan, Andrew C. Cullen, Simon R. Clarke}
		%\address{School of Mathematical Science, Monash University, Vic, 3800, Australia}
		
		\author[address1]{Keith C.H. Chan\corref{mycorrespondingauthor}}
		\cortext[mycorrespondingauthor]{Corresponding author}
		\ead{keith.chan@monash.edu}
		
		\author[address2]{Andrew C. Cullen}
		\author[address1]{Simon R. Clarke}

		\address[address1]{School of Mathematical Science, Monash University, Vic, 3800, Australia}
		\address[address2]{School of Computing and Information Systems, The University of Melbourne, Vic, 3010, Australia}

		\begin{abstract}
		The forced Korteweg-de Vries (fKdV) equation describes incompressible inviscid free surface flows over some arbitrary topography. We investigate solitary and hydraulic fall solutions to the fKdV equation. Numerical results show that the calculation of exponentially decaying solitary waves at the critical Froude number is a nonlinear eigenvalue problem. Furthermore we show how exponential decaying solitary waves evolve into the continuous spectrum of algebraic decaying solitary waves.
	
		A novel and stable numerical approach using the wave-resistance coefficient and tabletop solutions is used to generate the hydraulic fall parametric space. We show how hydraulic fall solutions periodically evolve into exponential decaying solitary waves.
		\end{abstract}
		
		
	\end{frontmatter}
	
	\linenumbers

	\section{Introduction}
	\label{section:Introduction}

	The forced KdV equation (fKdV) is an extension of the Korteweg-de Vries (KdV) equation that describes incompressible inviscid free surface flows in a two-dimensional channel over some arbitrary topography. The flow is characterised by the dimensionless flow rate given by the Froude number that is, the ratio between the flow velocity and wave velocity:
	\begin{equation}
	F=\frac{U}{\sqrt{gH}},
	\end{equation}
	where $g$ is the acceleration due to gravity, $H$ is the undisturbed fluid depth and $U$ is the flow velocity. The Froude number classifies the solution to the steady fKdV equation according to 11 basic steady flow types \cite{binder2019steady}. When $F<1$ the flow is classified as subcritical, and supercritical for $F>1$.\\

	We focus on steady hydraulic fall solutions which involve a transition between subcritical and supercritical flow. Hydraulic flow has been observed for various types of topographical disturbances including semi-circular (both numerically and experimentally by Forbes \cite{forbes1988critical}), bell-shaped \cite{dias2002generalised}, triangular \cite{dias1989open}, and delta function forcing \cite{shen1991locally}. We are especially interested in the relationship for hydraulic fall solutions between the two parameters $\Delta\propto F-1$ and topography height $\gamma$.\\
	
	Asymptotic methods can be used to determine the hydraulic fall parametric space in various limits. Grimshaw and Smyth \cite{grimshaw1986resonant} used asymptotic analysis to show the parametric space in the small forcing limit $\gamma\rightarrow \pm0$. In the large positive forcing limit where dispersive effects are negligible, hydraulic approximation can be used to give the relation $\Delta\propto\sqrt{\gamma}$. However, no known theoretical results exist in the negative forcing limit.\\
	
	Ee and Clarke \cite{ee2007weakly} numerically determined the parametric relationship for hydraulic fall solutions under a $\hbox{sech}^2$ forcing. Their results were consistent with the theoretical asymptotic analysis. A minimisation algorithm based on the Hamiltonian of the fKdV equation was used to obtain hydraulic fall solutions. An identical topography added far downstream allows asymmetric hydraulic fall solutions to be analysed in a symmetric manner giving tabletop solutions. Tabletop solutions are symmetric over the domain but are not symmetric over each bump. The flow transitions from a subcritical flow to supercritical flow over the first bump, then transitions back from supercritical flow to subcritical flow over the second bump. Despite one half of the tabletop solution domain being potentially unstable, tabletop solutions appear to be the symmetric variant of asymmetric hydraulic fall solutions. Asymmetric hydraulic fall solutions are then recovered by only considering half the domain of the symmetric tabletop solutions. Ee and Clarke observed that hydraulic fall solutions periodically transform into solitary waves in the negative $\gamma$ region. Two solitary wave solutions were embedded in this parametric relationship. As the topographic forcing is exponentially decaying, these solitary waves therefore have exponential far field decay. We will refer to them as Exponential Decaying Solitary Waves (EDSW). Tabletop solutions have been similarly observed by Chardard et al \cite{chardard2011stability} with a $\hbox{sech}^2$ forcing and Lee and Whang \cite{lee2015trapped} with a $\cos^4$ forcing.\\
	
	Cullen \cite{cullen2018novel} similarly mapped a portion of the hydraulic parametric space. A minimisation of the wave-resistance coefficient \cite{wu1987generation} was instead used to filter for hydraulic fall solutions, with artificial boundary conditions for the truncated numerical domain. However, both Ee and Clarke, and Cullen experienced numerical difficulties in mapping the extended negative $\gamma$ region.\\
	
	Another type of solitary wave was observed by Keeler et al \cite{keeler2017critical} (henceforth referred to as KBB). These solitary wave solutions exist on a continuum in $\gamma$. Asymptotic analysis show that these solutions have algebraic far field decay. Furthermore numerical results show multiple types of algebraically decaying solitary solutions not predicted by their asymptotic analysis. We will refer to these as Algebraic Decaying Solitary Waves (ADSW).\\
	
	The goal of this paper is to firstly develop a stable numerical method in mapping the extended negative region in the hydraulic parameter space. Secondly to explore the connection between KBB's continuum of ADSW solutions to the discrete EDSW found by Cullen. Finally, to determine the relationship between hydraulic fall solutions and EDSW solutions.
	
	\section{Problem formulation}
	\label{section:Problem formulation}
	
	The fKdV equation in dimensionless form is
	\begin{equation}
	A_t-\Delta A_x+6AA_x+A_{xxx}=-\gamma f_x,
	\label{fkdv}
	\end{equation}
	where $A$ is the wave amplitude, $x$ and $t$ are length and time variables respectively, $f$ is the external topographic forcing, $\Delta$ is the detuning parameter and $\gamma$ is the forcing coefficient. For free surface flow, $A$ corresponds to the free surface displacement, scaled by the undisturbed fluid depth $H$, $x=\left(3/2\right)^{1/2}x'/H$, $t=\left(3/2\right)^{3/2}t'/(H/g)^{1/2}$, $\Delta=4(F-1)$ and $\gamma=2$, where $x'$ and $t'$ are respectively the dimensional horizontal space and time variables.\\
	
	For steady flow, equation \eqref{fkdv} can be integrated once in $x$ to obtain
	\begin{equation}
	-\Delta A+3A^2+A_{xx}=-\gamma f,
	\label{sfkdv}
	\end{equation}
	as considered by Ee and Clarke \cite{ee2007weakly}. The downstream boundary conditions, which eliminates the constant of integration, are given by
	\begin{equation}
	A\rightarrow0,\quad A_x\rightarrow0,\quad f\rightarrow 0 \quad\text{as}\quad x\rightarrow-\infty.
	\label{bc}
	\end{equation}
	
	We follow Ee and Clarke and use equation \eqref{sfkdv}, with $f=\hbox{sech}^2x$ as the chosen bump forcing. There are several reasons for this choice. Firstly, this allows comparison to the numerical results by Ee and Clarke. Moreover, several analytic solitary wave solutions are available, which provides a natural starting point for the numerical continuation process. Additionally, perturbation analysis can be performed around known analytic solutions. Throughout, we focus on hydraulic fall and solitary wave solutions to equation \eqref{sfkdv}.\\
	
	The Hamiltonian for the steady fKdV equation is given by
	\begin{equation}
	H(A,A_x)=\gamma fA-\frac{1}{2}\Delta A^2+\frac{1}{2}A_x^2+A^3.
	\label{hamiltonian}
	\end{equation}
	An understanding of hydraulic fall solution dynamics can be found by observing the Hamiltonian in the far field where topographic effects are negligible; that is $f=0$. The phase portrait is then given by 
	\begin{equation}
		A_x^2=\Delta A^2-2A^3+C.
	\end{equation}
	The phase plane has two critical points. In the case $\Delta>0$, the critical points are a saddle point at $(A,A_x)=(0,0)$ and a centre (minimum) at $(A,A_x)=(\Delta/3,0)$. Hydraulic fall solutions traverse from one critical point to the other in the phase plane. The $\Delta<0$ case simply reverses the nature of the critical points where $(A,A_x)=(0,0)$ becomes the saddle point while $(A,A_x)=(\Delta/3,0)$ becomes the centre. Converting from $\Delta>0$ to the $\Delta<0$ case can be made simply through the transformation $A\leftarrow A-\Delta/3$. Thus without loss of generality we assume $\Delta>0$. Similarly, without loss of generality, we assume the downstream limit as the saddle point since the flow is reversible.\\
	
	In the special case of critical flow when $\Delta=0$, the critical points coincide and a solitary wave solution path traverse from the origin $(A,A_x)=(0,0)$, back to the origin.\\
	
	\subsection{Solitary wave solutions}
	\label{subsection:Solitary wave solutions}
	
	Solitary waves are solutions to the fKdV equation in the critical flow case where $\Delta=0$, with the boundary conditions $A\rightarrow0$ as $x\rightarrow\pm\infty$ given by the equation
	
	\begin{equation}
		A_{xx}+3A^2=-\gamma \hbox{sech}^2(x).
		\label{criticalfKdV}
	\end{equation}
	
	KBB showed the condition $\gamma\int_{-\infty}^{\infty}f(x)dx\le 0$ is necessary for the existence of a solitary wave solution. This can be obtained by integrating \eqref{criticalfKdV} and applying the far field boundary conditions \eqref{bc}. Therefore solitary wave solutions only exist for negative $\gamma$ for our chosen bump forcing.\\
	
	Topographical effects for our chosen bump forcing become negligible as $x\rightarrow\pm\infty$. Solitary wave solutions far upstream and downstream must satisfy the equation
	\begin{equation}
	3A^2+A_{xx}=0.
	\label{predictdecay}
	\end{equation}
	This has solutions
	\begin{equation}
	A=c/x^2
	\label{algebraicdecay}
	\end{equation}
	for $c=0,-2$, corresponding to EDSW and ADSW. ADSW solutions were found by KBB using Gaussian forcing and the scaled equation in the form $u_{\xi\xi}+u^2=\alpha e^{-\xi^2}$. Matched asymptotics were used for both $\alpha\ll1$ and $\alpha\gg1$, confirming algebraic far field decay. Numerical results by KBB confirmed the predicted matched asymptotic behaviour. However, KBB also found several extra types of numerical solutions not predicted by their asymptotic analysis. Furthermore, there appears to be an infinite sequence of new solutions with oscillations localised around the origin.\\
	
	Analytic EDSW solutions are also known for equation \eqref{criticalfKdV}. These can be found by assuming a solution in the form $A=\alpha \hbox{sech}^2(x)$ for some unknown constant $\alpha$ \cite{camassa1991stability}. Solving for the constant $\alpha$ gives the zero solution $A=0$ at $\gamma=0$ and a EDSW solution $A=2\hbox{sech}^2(x)$ at $\gamma=-8$.\\
	
\begin{figure}[H]
	\centering
	\begin{subfigure}[b]{0.49\textwidth}
		\includegraphics[width=\textwidth]{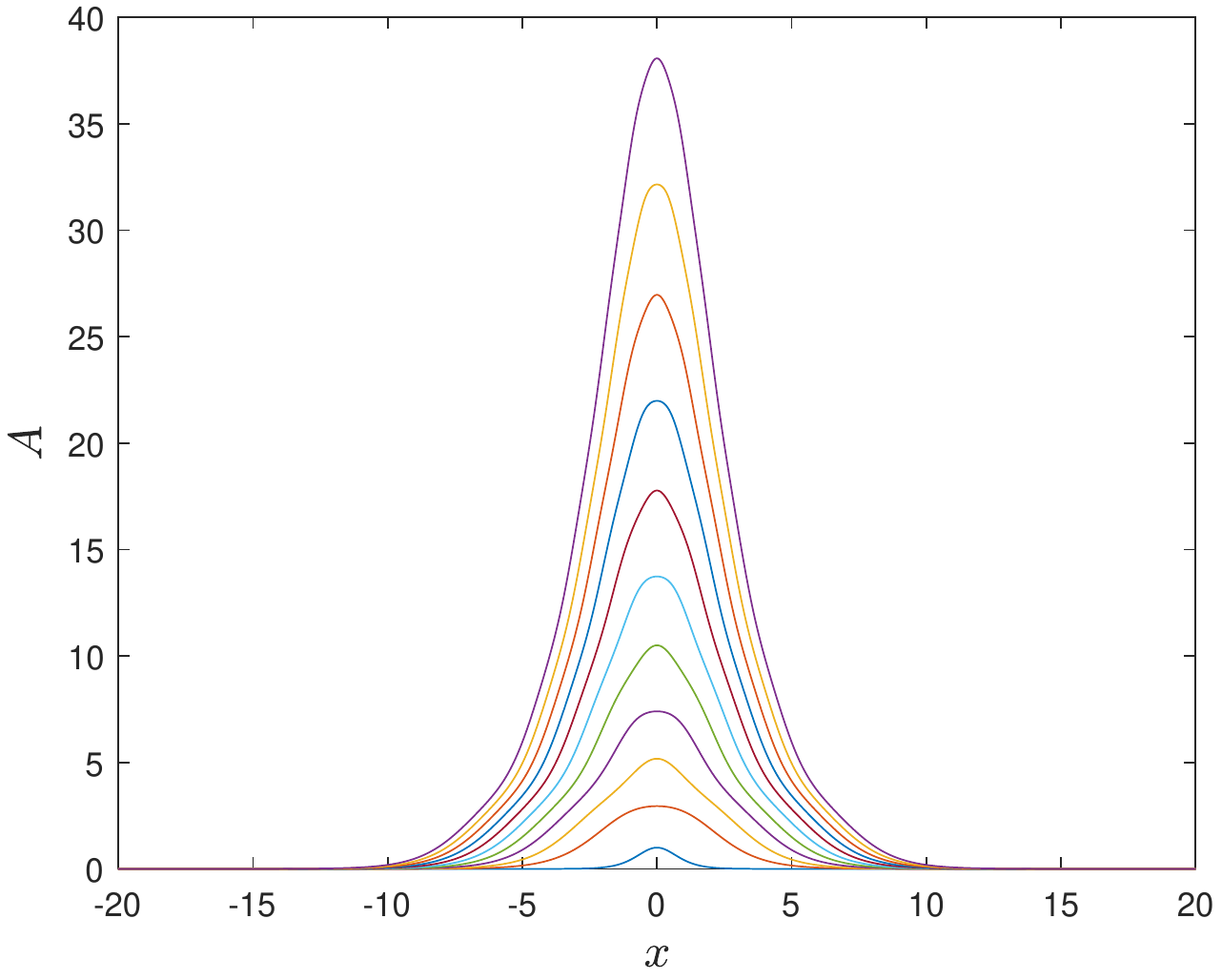}
	\end{subfigure}
	\begin{subfigure}[b]{0.49\textwidth}
		\includegraphics[width=\textwidth]{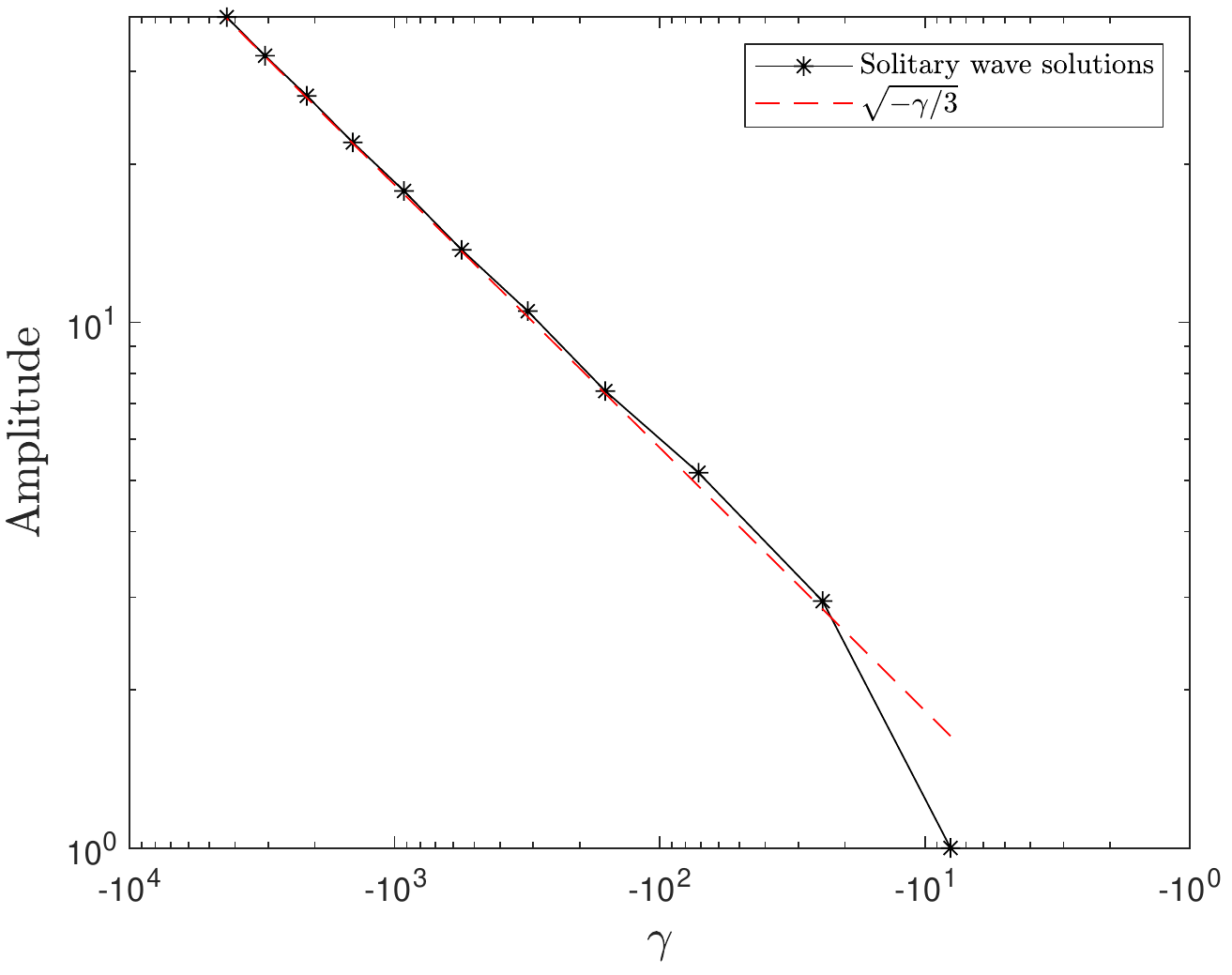}
	\end{subfigure}
	\caption{First 11 EDSW solutions to equation \eqref{criticalfKdV} obtained using a shooting method (left). EDSW amplitude vs $\gamma$ for the solutions obtained and comparison to the large $\gamma$ limit $|A|\sim\sqrt{\frac{-\gamma}{3}}$ (right).}
	\label{solitary}
	\end{figure}

	To investigate further EDSW solutions, a shooting method was used with initial conditions $(A,A_x)=(0,0)$ as $x\rightarrow\infty$ for a range of $\gamma$ values. We apply a secant method for the necessary condition $A_x(0)=0$ for solitary wave solutions. The first solitary wave found matches the EDSW analytic solution at $\gamma=-8$, while the second solitary wave at $\gamma\simeq-24.3$ agrees with Ee and Clarke to within an error of $1\%$. Several more solitary wave solutions were found at various $\gamma$ values shown in Figure \ref{solitary}. Hence, equation \eqref{criticalfKdV} corresponds to a nonlinear eigenvalue problem, with eigenvalues $\gamma$ and EDSW eigenfunctions. The first 11 eigenvalues are shown in Table \ref{table1}, with the location of the eigenvalues following a power-law relation numerically found to be $\gamma\sim-1.23\times n^{3.4}$ for $n\gg1$. In the limit as $\gamma\rightarrow-\infty$, following KBB's matched asymptotic approach to determine the solution behaviour in the vicinity of the topography $|x|\ll1$, we obtain the asymptotic behaviour 
	\begin{equation}
	|A|\sim\sqrt{\frac{-\gamma}{3}}.
	\label{asymptoteA}
	\end{equation}
	Figure \ref{solitary} shows agreement with the predicted large $\gamma$ behaviour.
	
	\begin{table}[h]
	\begin{subtable}[h]{\textwidth}
		\centering
		\begin{tabularx}{1\textwidth} { |X|X|X|X|X|X|X|}
			\hline
			n & 1 & 2 & 3 & 4 & 5 & 6\\
			\hline
			$\gamma$  & -8 & -24.29 & -71.42 & -160.44 & -314.51 & -558.37\\
			\hline
		\end{tabularx}
	\end{subtable}
	\vspace{0.5cm}
	\newline
	\begin{subtable}[h]{\textwidth}
		\centering
		\begin{tabularx}{1\textwidth} { |X|X|X|X|X|X|X|}
			\hline
			n & 7 & 8 & 9 & 10 & 11 \\
			\hline
			$\gamma$ & -921.42 & -1437.12 & -2140.18 & -3080.82 &-4296.34  \\
			\hline
		\end{tabularx}	
	\end{subtable}
	\caption{First 11 eigenvalues $\gamma$ to 2 decimal places, corresponding to EDSW eigenfunctions to equation \eqref{criticalfKdV}.}
	\label{table1}
	\end{table}

	These numerical results suggest there may be an infinite number of eigenvalues and corresponding EDSW eigenfunctions to \eqref{criticalfKdV}.

	\subsection{Hydraulic fall solutions}
	\label{subsection:Hydraulic fall solutions}
	
	Hydraulic fall solutions are solutions in the general case $\Delta\ne0$, and traverse between the two critical points in the phase plane. We begin by exploring analytic perturbation and asymptotic results.
	
	\subsubsection{Analytic results}
	\label{subsubsection:Analytic results}
	
	Asymptotic analysis of the parameter space $\gamma(\Delta)$ for hydraulic fall solutions in the limit $\gamma\rightarrow 0$ was shown by Grimshaw and Smyth \cite{grimshaw1986resonant} to give the relationship
	
	\begin{equation}
		\left(\frac{\Delta}{3}\right)^3=(K\gamma)^2
		\label{smallgamma}
	\end{equation}
	with $K=\int_{-\infty}^{\infty}fdx$. In the limit as $\gamma\rightarrow\infty$, the dispersive term in equation \eqref{sfkdv} becomes negligible. This results in a quadratic equation in $A$, where the parametric relationship for large positive $\gamma$ matched at the topographical maximum is given by
	\begin{equation}
		\Delta=\sqrt{12\gamma}.
		\label{largegamma}
	\end{equation}
	To our knowledge, there are no known asymptotic limits for $\gamma\rightarrow-\infty$.\\
	
	We turn to perturbation methods to find the parametric relationship $\Delta(\gamma)$ in the vicinity of the analytic solution at $\gamma=-8$. We use a regular perturbation series in $A$ and $\gamma$ for
	
	\begin{equation}
	-\epsilon A+3A^2+A_{xx}=-\gamma f,
	\label{perturbationequation}
	\end{equation}
	and apply the hydraulic fall boundary conditions from the phase plane
	
	\begin{equation}
	A\rightarrow 0 \quad\text{as}\quad x\rightarrow-\infty \quad\text{and}\quad A\rightarrow \Delta/3 \quad\text{as}\quad x\rightarrow\infty.
	\label{hydraulicbcs}
	\end{equation}
	We find in the vicinity of $\gamma=-8$ for $\Delta\ll1$, the parametric curve for hydraulic solutions to order $\epsilon^2$ must lie on the curve
	\begin{equation}
	\gamma=-8+\frac{31}{72}\Delta^2+O(\Delta^4).
	\end{equation}
	Similarly, solitary wave solutions can be found by applying the zero boundary condition for the upstream and downstream height to equation \eqref{perturbationequation}. This gives the solitary wave solution in the form $A=2\hbox{sech}^2(x)$ on the line $\Delta=\frac{1}{2}(\gamma+8)$.\\
	
	The perturbation analysis agrees with the numerical results obtained by Ee and Clarke but is unable to obtain the negative branch emanating from $\gamma=-8$. We resort to numerical results for the extended negative region.\\
	
	\subsubsection{Numerical results}
	\label{subsubsection:Numerical results}
	
	The fKdV equation consist of two free parameters and requires an additional condition for a unique solution. Ee and Clarke \cite{ee2007weakly} and Cullen \cite{cullen2018novel} both utilised a minimisation algorithm based on the Hamiltonian, to locate hydraulic fall solutions. Cullen used a minimisation of the wave-resistance coefficient introduced by Wu \cite{wu1987generation}. By imposing the upstream and downstream boundary conditions for hydraulic fall solutions; that is $H(0,0)=0$ and $H(\Delta/3,0)=\Delta^3/54$, the net change in Hamiltonian between $x=\pm \infty$ is calculated to be
	\begin{equation}
	\delta H=\Delta^3/54.
	\end{equation}
	Differentiating and then integrating the Hamiltonian over the real line, we obtain the wave-resistance coefficient which must equal the net change in Hamiltonian given by
	\begin{equation}
	\int_{-\infty}^{\infty}\frac{dH}{dx}dx=\int_{-\infty}^{\infty}\gamma f_x Adx\le \frac{\Delta^3}{54}.
	\label{wave-resistance}
	\end{equation}
	
	The parametric space $\Delta(\gamma)$ for hydraulic fall solutions was numerically determined by Ee and Clarke\cite{ee2007weakly}. A shooting method in combination with a minimisation algorithm was used to locate solutions which converged to the downstream saddle point. This was built upon a consideration of the Hamiltonian. The conditions imposed ensure the downstream amplitude lies within the homoclinic orbit of the phase plane. The minimisation conditions used in their algorithm are
	\begin{equation}
		\begin{split}
			H(H-H_s)\le 0 \\
			-\frac{\Delta}{3}\le A \le \frac{\Delta}{6}
		\end{split}
	\end{equation}
	where $H_s=H(\Delta/3,0)=\Delta^3/54$. Ee and Clarke utilised tabletop solutions, allowing the use of Fourier spectral methods.\\
	
	Ee and Clarke's results show that the hydraulic fall solutions evolve into the analytic solitary wave solution at $\gamma=-8$. Furthermore, a second solitary wave solution was found numerically at $\gamma\simeq -24.55$. Between the zero solution at $\gamma=0$ and the solitary wave at $\gamma=-8$, and the two subsequent solitary waves at $\gamma=-8$ and $\gamma\simeq 24.55$, the connecting parametric curve followed an arc shape. The hydraulic fall solutions and solitary wave solutions in the continuation exhibit exponential far field decay. The EDSW solutions discussed in Section~\ref{subsection:Solitary wave solutions} also coincide with the $\Delta=0$ values in the hydraulic fall solution. Both the small $\gamma$ approximation \eqref{smallgamma} and large positive $\gamma$ limit \eqref{largegamma} are consistent with the numerical results of Ee and Clarke.\\
	
	Cullen showed asymmetric solutions must correspond to a minima of the wave-resistance coefficient \eqref{wave-resistance}, with equality for hydraulic fall solutions. A Chebyshev collocation method as detailed by Cullen and Clarke \cite{cullen2019fast} was used for a truncated numerical domain with artificial boundary conditions simulating zero far field boundary conditions. This method potentially introduces nonlinear waves in the downstream limit. To combat this, a minimisation process was chosen over the integral condition. A genetic algorithm was used for the minimisation criteria and was able to replicate the parametric relationship obtained by Ee and Clarke for the region $-10\le\gamma\le 5$, however numerical difficulties occurred for $\gamma\le -8$.\\
	
	We aim to extend the hydraulic fall parametric relationship results by Ee and Clarke, and Cullen. Their results suggest that between each subsequent EDSW solutions in Figure \ref{solitary} there may also exist some connecting arc-shaped parametric relationship $\gamma(\Delta)$. The numerically obtained EDSW solutions from the shooting method and their corresponding $\gamma$ value locations can then be used as a natural starting point for the continuation process.\\
	
	Both Ee and Clarke, and Cullen encountered numerical difficulties in mapping the extended negative region. We develop several improvements for numerical stability. Firstly, we use the wave-resistance constraint which hydraulic fall solutions must satisfy. Secondly, similar to the approach by Ee and Clarke, a second identical bump forcing is added downstream which converts the asymmetric problem into a symmetric one. Boundary conditions are automatically enforced using Fourier spectral methods.
		
	\section{Numerical Method}
	\label{section:Numerical Method}
	
	To obtain tabletop solutions of the steady fKdV equation, we consider forcing consisting of two identical bumps separated by distance $2h$, given by $f(x)=\hbox{sech}(x-h)^2+\hbox{sech}(x+h)^2$. Since the wave-drag constraint applies to asymmetric hydraulic fall solutions, only half of the domain is used. We look for the solution $A(\gamma,\Delta)$ and the hydraulic parametric relationship $\Delta(\gamma)$ to the nonlinear system
	\begin{equation}
	A_{xx}+3A^2-\Delta A=-\gamma f(x)
	\label{sfkdv2}
	\end{equation}
	\begin{equation}
	\int_{0}^{\infty}\gamma f_x Adx=\int_{-\infty}^{0}\gamma f_x Adx=\frac{\Delta^3}{54}
	\label{wavedrag}
	\end{equation}
	on $x\in(-\infty,\infty)$ with the hydraulic fall boundary conditions \eqref{hydraulicbcs}. The problem is not strictly symmetric but can be treated symmetrically on a sufficiently large numerical domain. Newton's method is used to solve the system of nonlinear equations with the Jacobian of equation \eqref{sfkdv2} given by
	\begin{equation}
		J(v)=v_{xx}+(\Delta+6A)v.
		\label{jacobian}
	\end{equation}
	
	\subsection{Fourier Collocation methods}
	\label{subsection:Fourier Collocation methods}
	We discretise the Jacobian \eqref{jacobian} within each Newton iteration using Fourier collocation methods. This requires us to solve the linear system $\Lambda\textbf{u}=\textbf{f}$ at each Newton step. The Fourier collocation points on the finite periodic domain $(-L/2,L/2)$, are given by $N$ evenly spaced points
	\begin{equation}
		x_j=\frac{L j}{N}-\frac{L}{2},\quad j=0,1,\dots,N-1.
	\end{equation}

	Gaussian quadrature is used to approximate integrals such as the wave-resistance \eqref{wavedrag}
	\begin{equation}
		\int_{0}^{L/2}f(x)dx\simeq\sum_{j=0}^{N/2}w_jf(x_j).
	\end{equation}
	In the case of a periodic symmetric function, spectral accuracy can be obtained using weights identical to the trapezoidal rule; that is $w_0=w_{N/2}=\Delta x/2$ and $w_j=\Delta x$ at the Fourier collocation points.\\
	
	Fourier differentiation matrices can be used to approximate derivatives. However, explicitly using the differentiation matrices requires $O(N^2)$ operations for the matrix vector multiplication. An alternative approach is to perform differentiation using Fourier transforms. Let $u_j$ be the approximation to $u(x_j)$ at the collocation points $x_j$, and $p$ be the wave numbers associated with the Fourier modes. Using the Discrete Fourier Transform (DFT), the Fourier coefficients of $u_j$ are then given by
	\begin{equation}
		{u}_j=\frac{1}{N}\sum_{p=-N/2}^{N/2-1}\hat{u}_pe^{ipx_j},\quad p=-N/2,-N/2+1,\dots,N/2-1.
	\end{equation}
	In spectral space the derivative of $\hat{u}_j$ is simply given by
	\begin{equation}
		du/dx(x_j)=\frac{1}{N}\sum_{p=-N/2+1}^{N/2-1}ip\hat{u}_pe^{ipx_j}.
	\end{equation}
	Given $u_j$ is real, the mode $j=-N/2$ in the derivative contains a purely imaginary component and thus is filtered \cite{zang1984spectral}. Only odd derivatives require the filtering of the highest mode. Let $C$ be the matrix representing the DFT matrix with elements
	\begin{equation}
		C_{pj}=\frac{1}{N}e^{-ipx_j}
	\end{equation}
	and $C^{-1}$ be the inverse DFT matrix. We define the diagonal matrix $\tilde{D}$ with diagonals $ip$ as the differentiation matrix in Fourier space 
	\begin{equation}
		\tilde{D}_{pq}^n=(ip)^n\delta_{pq}
	\end{equation}
	where $n$ is the order of the derivative and $\delta$ is the Kronecker delta. Thus the $n$th spectral differentiation matrix can be represented as
	\begin{equation}
		D^n=C\tilde{D}^nC^{-1}.
	\end{equation}

	Using this approach, matrix multiplication can be efficiently performed using the Fast Fourier Transform \cite{cooley1965algorithm} (FFT) in $O(N\log N)$ operations even though the spectral differentiation matrix is full. A further benefit is less memory usage since the spectral matrix $\Lambda$ is not explicitly saved, but simply defined as an operator consisting of Fourier transforms. Spectral multigrid can then be used to solve the Jacobian discretisation.\\
	
	\subsection{Spectral Multigrid}
	\label{subsection:Spectral Multigrid}
	
	Spectral Multigrid (SMG) was initially proposed by Zang et al \cite{zang1982spectral,zang1984spectral} and combines elements of spectral methods with multigrid concepts. The main advantage of spectral approximation is the ability to achieve accurate results with fewer grid points than typical finite difference approximation due to their spectral or infinite order convergence. Fourier spectral methods are used for periodic problems. This allows the use of the FFT reducing the number of operations required for a solution from $O(N^2)$ to $O(N\log N)$.\\
	
	\subsubsection{Restriction and Prolongation operators}
	\label{subsubsection:Restriction and Prolongation operators}
	
	Grid transfer operators used in multigrid should have comparable accuracy to the discretisation scheme. For traditional multigrid with finite difference, polynomial interpolation is commonly used. Fourier SMG utilises trigonometric interpolation, which can be efficiently performed using the FFT. The restriction operator simply consists of taking the Fourier transform, removing the upper half frequencies, then taking the inverse Fourier transform. Similarly, the prolongation operator involves taking the Fourier transform, padding the upper half frequencies with zeros, then taking the inverse Fourier transform. This allows the restriction and prolongation operators to be computed using two FFTs.
	
	\subsubsection{Relaxation}
	\label{subsubsection:Relaxation}
		
	Relaxation choices are also restricted since the spectral differentiation matrices are dense. Therefore, commonly used relaxations for multigrid such as Gauss-Seidel cannot be used efficiently. Instead, relaxations that update all values at once are required allowing the use of FFT to perform the relaxations. Richardson type iterations are preferred over Jacobi since the latter requires knowledge of the diagonal elements of the spectral matrix. Since we have not defined the linear operator $\Lambda$ explicitly but instead through Fourier operators, the diagonal elements are not readily known.\\
	
	Richardson iterations to the linear system $\Lambda\textbf{u}=\textbf{f}$ are of the form
	
	\begin{equation}
		\textbf{v}^{(k+1)}=\textbf{v}^{(k)}+\omega \textbf{r}^{(k)},
	\end{equation}
	where $\textbf{v}$ is the best estimate to the exact solution $\textbf{u}$, $\omega$ a chosen parameter, $\textbf{r}$ the residual and $k$ the iteration number.\\
	
	A parameter free iterative scheme introduced by Zang \cite{zang1982spectral} is Minimum Residual Richardson (MRR). MRR has the advantage of not requiring the knowledge of the minima and maximal eigenvalues of the spectral matrix but requires that the eigenvalues satisfy $Re(z)>0$ in the complex plane \cite{boyd2001chebyshev}. This is ensured with the use of a simple second-order finite difference preconditioner. The choice of relaxation parameter in MRR is to minimize the residual with 
	
	\begin{equation}
		\omega=\frac{(\textbf{r}^{(k)},\Lambda\textbf{v}^{(k)})}{(\Lambda\textbf{v}^{(k)},\Lambda\textbf{v}^{(k)})}
	\end{equation}
	where $(,)$ denotes the inner product.
	
	\subsection{Numerical continuation}
	\label{subsection:Numerical continuation}
	
	Numerical continuation methods are required to generate the hydraulic fall parametric space $\gamma(\Delta)$. While Ee and Clarke, and Cullen used a simple natural parameter continuation, we employ a pseudo arc-length continuation.\\
	
	Let $[F,G]$ be the approximation to \eqref{sfkdv2} and \eqref{wavedrag} respectively at the Fourier collocation points and $i$ the index at each numerical continuation step. Then the resulting system is given by
	\begin{equation}
		F=(A_{xx})_{j}^{(i)}+3A_{j}^{2{(i)}}-\Delta^{(i)} A_{j}^{(i)}+\gamma^{(i)} f(x_j),
	\end{equation}
	\begin{equation}
		G=\sum_{i=0}^{N/2} w_j\gamma^{(i)} f_x(x_j) A_{j}^{(i)},
	\end{equation}
	where $\Delta^{(i)}=\Delta(\gamma^{(i)})$ and $A_{j}^{(i)}=A(x_j,\gamma^{(i)},\Delta(\gamma^{(i)}))$. Let $\boldsymbol{\lambda}^{(i)}$ be the parameter vector $[\gamma^{(i)},\Delta^{(i)}]$ at each continuation step. We solve the system $[F,G]=0$ using pseudo arc-length continuation  with Euler predictor and Newton corrector \cite{keller1977numerical}. This requires the solution to three systems of linear equations at each step of the Newton corrector.
	
	\section{Numerical Results}
	\label{section:Numerical Results}
	
	\subsection{Tabletop Solutions}
	\label{subsection:Tabletop Solutions}
	
\begin{figure}[H]
		\centering
		\begin{subfigure}[b]{0.5\textwidth}
			\includegraphics[width=\textwidth]{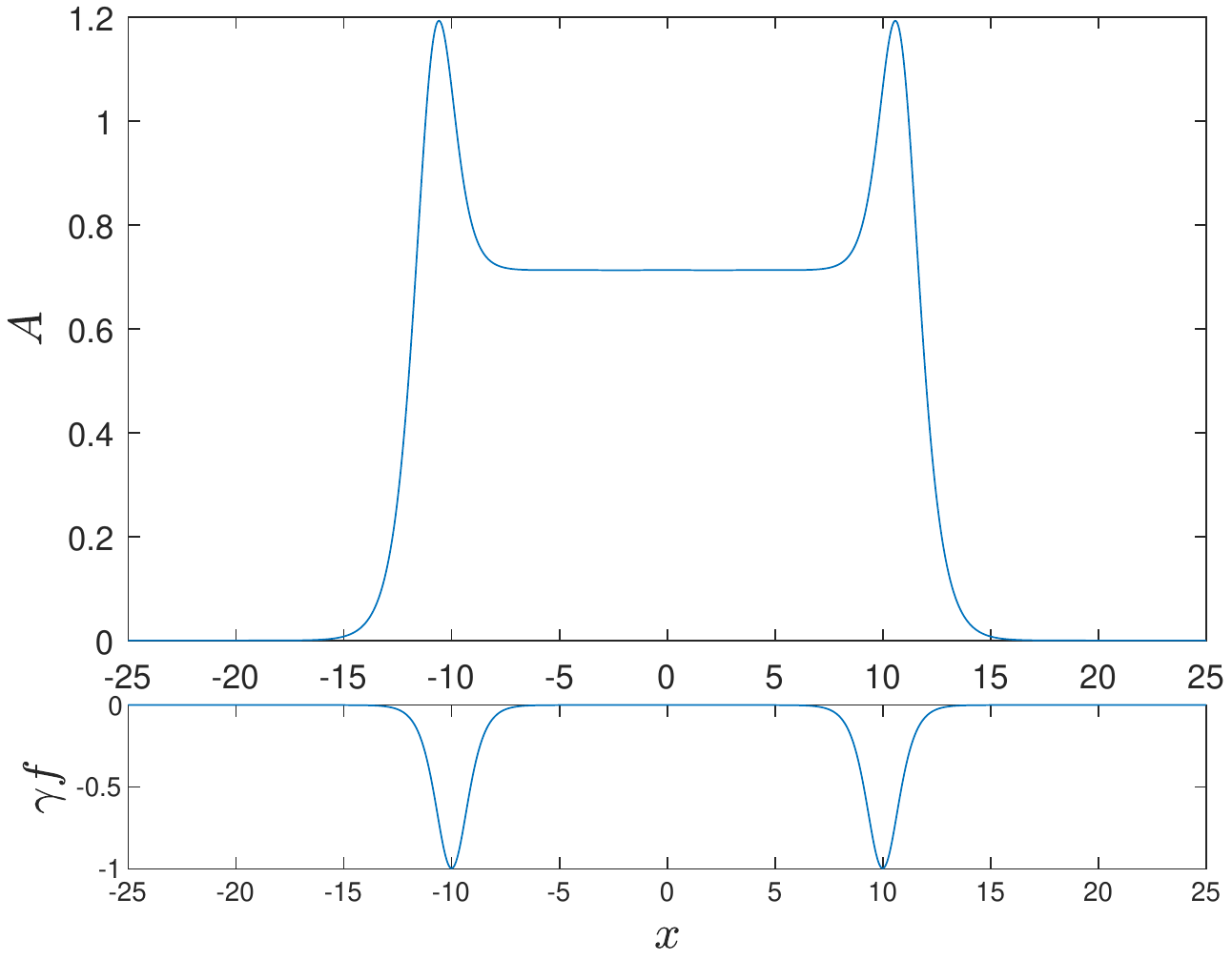}
		\end{subfigure}%
		\begin{subfigure}[b]{0.5\textwidth}
			\includegraphics[width=\textwidth]{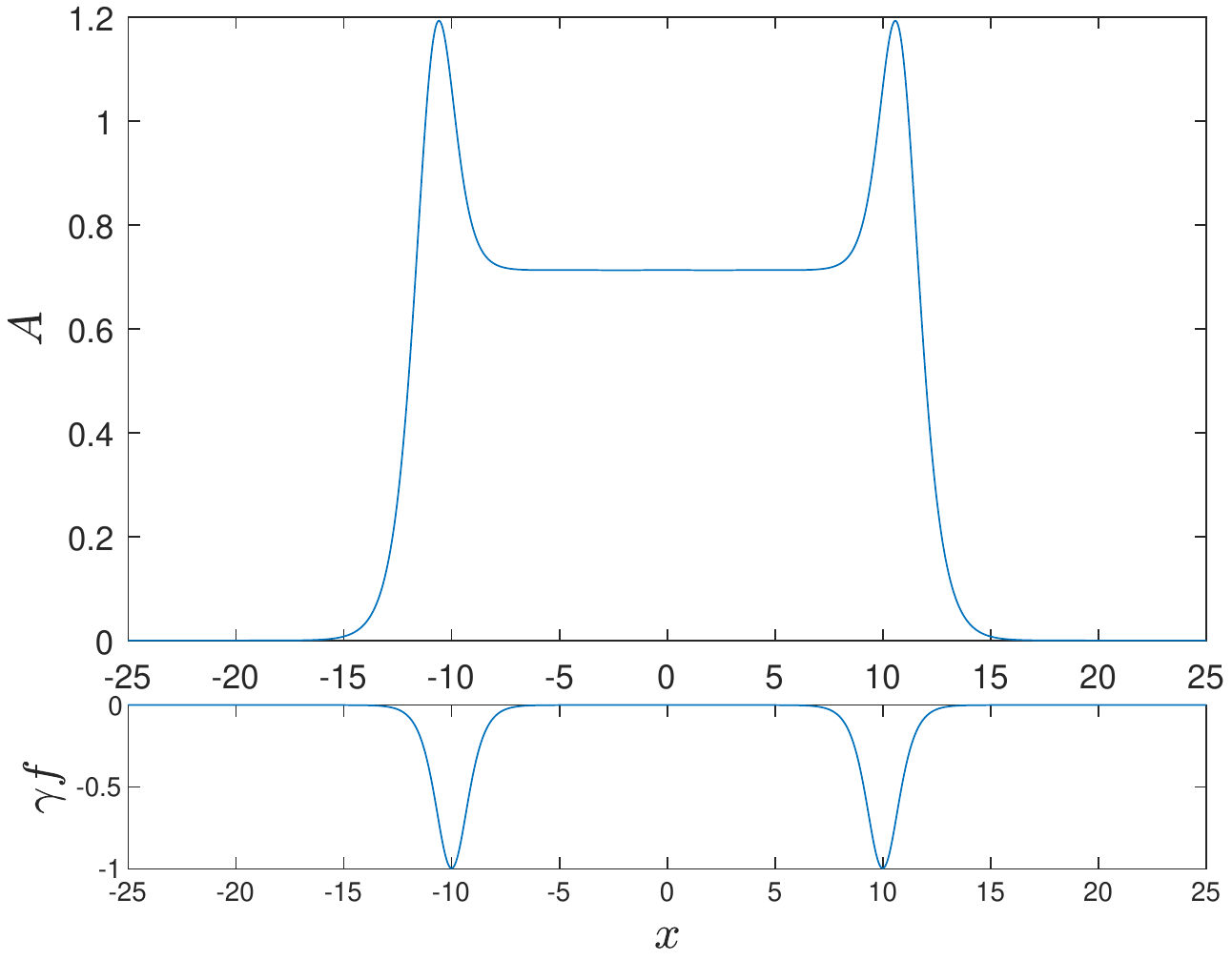}
		\end{subfigure}
		\caption{Tabletop solutions (upper) to equation \eqref{sfkdv} and topography (lower) for $\gamma=1$ (left) and $\gamma=-1$ (right)}
		\label{tabletop}
	\end{figure}

	The numerical results for tabletop solutions presented here utilise $2^{13}$ grid points, a domain of $L=200$ and bump spacing of $h=10$. It is important that $h$ is large enough so that the bumps are adequately separated. However, a large $h$ requires a larger numerical domain $L$ for the periodic assumption to hold. This in turn determines the minimum grid points needed to resolve the wave. Provided these conditions are met, the results are largely independent of the parameters $L$ and $h$. The coarsest grid in the multigrid algorithm was set at $2^5$ grid points. One V-cycle consisting of 1 MRR relaxation on the up and down cycle is used within each Newton iteration, with a residual tolerance set to $10^{-10}$. Examples of tabletop solutions obtained using SMG for positive and negative forcing are given in Figure \ref{tabletop}. This approach improves upon the methods used by both Ee and Clarke, and Cullen, since it eliminates the need for a minimisation algorithm. However care must be taken in the choice of numerical domain and the distance between bumps.\\
		
	Numerical results, which are supported by the results of Lee and Whang \cite{lee2015trapped} suggest that tabletop solutions only depend on the topography height and the Froude number and are independent of the distance between topographies $h$, provided $h$ is large enough to avoid overlap and trailing tail influence. This is to be expected if tabletop solutions are symmetric variants of asymmetric hydraulic fall solutions.
	
	\subsection{Critical flow solutions}
	\label{subsection:Critical flow solutions}
	
	Our numerical investigations suggest the ADSW solutions obtained by KBB originate from EDSW solutions. The reasoning is two-fold: Firstly, the branch termination point for different types of ADSW appear to coincide with the location of the EDSW. Secondly, during the continuation process used to find the hydraulic parametric relationship, after reaching a solitary wave eigenvalue, the solution would occasionally jump to a type of ADSW solution identical to that observed by KBB. Furthermore, the type of ADSW obtained at the jump depended on which eigenvalue was approached. For example, type I solutions obtained after the zero solution, type II after the first EDSW, type IIa after the second EDSW etc.\\
	
\begin{figure}[H]
		\centering
		\begin{subfigure}[b]{0.5\textwidth}
			\includegraphics[width=\textwidth]{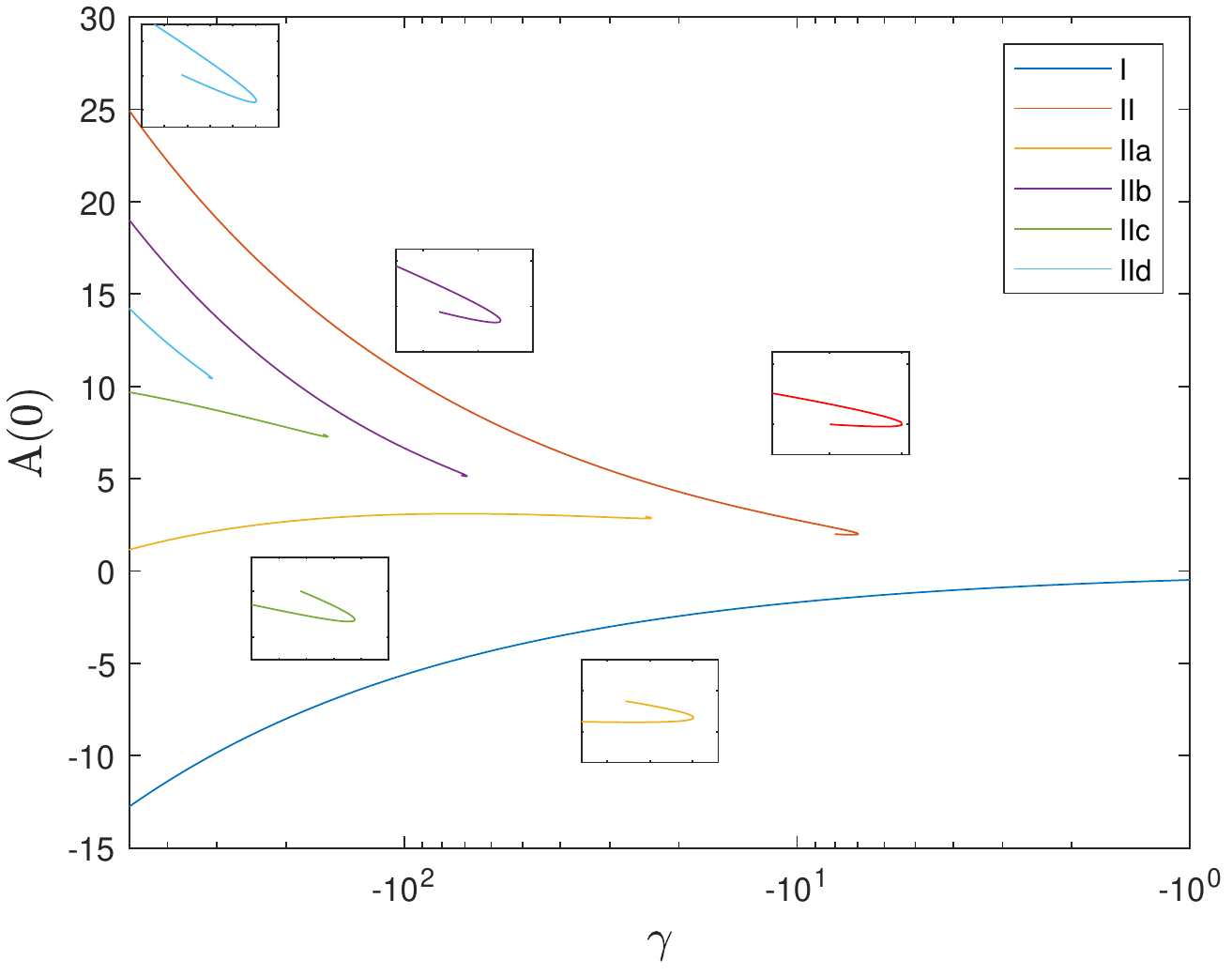}
		\end{subfigure}%
		\begin{subfigure}[b]{0.5\textwidth}
			\includegraphics[width=\textwidth]{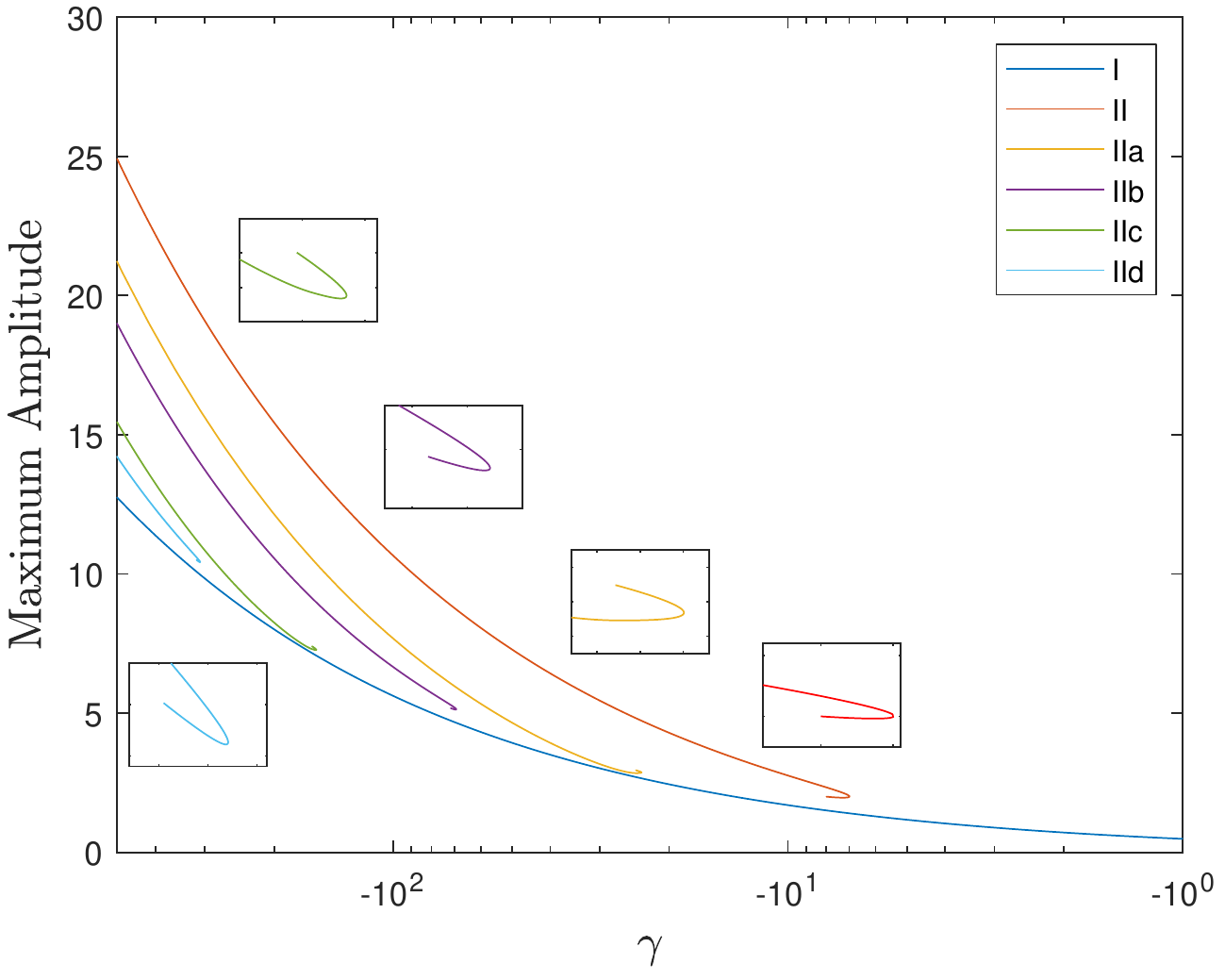}
		\end{subfigure}
		\caption{ADSW to equation \eqref{sfkdv2} similar to that obtained by KBB. The turning points are enlarged for visibility, capturing the transition from EDSW to ADSW. Plot labels are consistent with KBB for comparison.}
		\label{keeler}
	\end{figure}

	We test this hypothesis by using the numerically obtained EDSW from the shooting method as the starting point for the continuation process. The solution path obtained from the shooting method is traced using pseudo arc-length continuation with parameter $\gamma$. Solution types I, II, IIa, IIb, IIc, IId from KBB are replicated in Figure \ref{keeler} with each type originating from a different EDSW. A small turning point is observed briefly during the transition from EDSW to ADSW. Numerical results from our shooting method suggests there may be an infinite set of EDSW. If each EDSW also evolves into a separate type of ADSW, this suggest there may be infinitely many types of ADSW existing on $\Delta=0$, with a new solution type continuum appearing after each new EDSW. \\
	
\begin{figure}[H]
		\includegraphics[width=\textwidth]{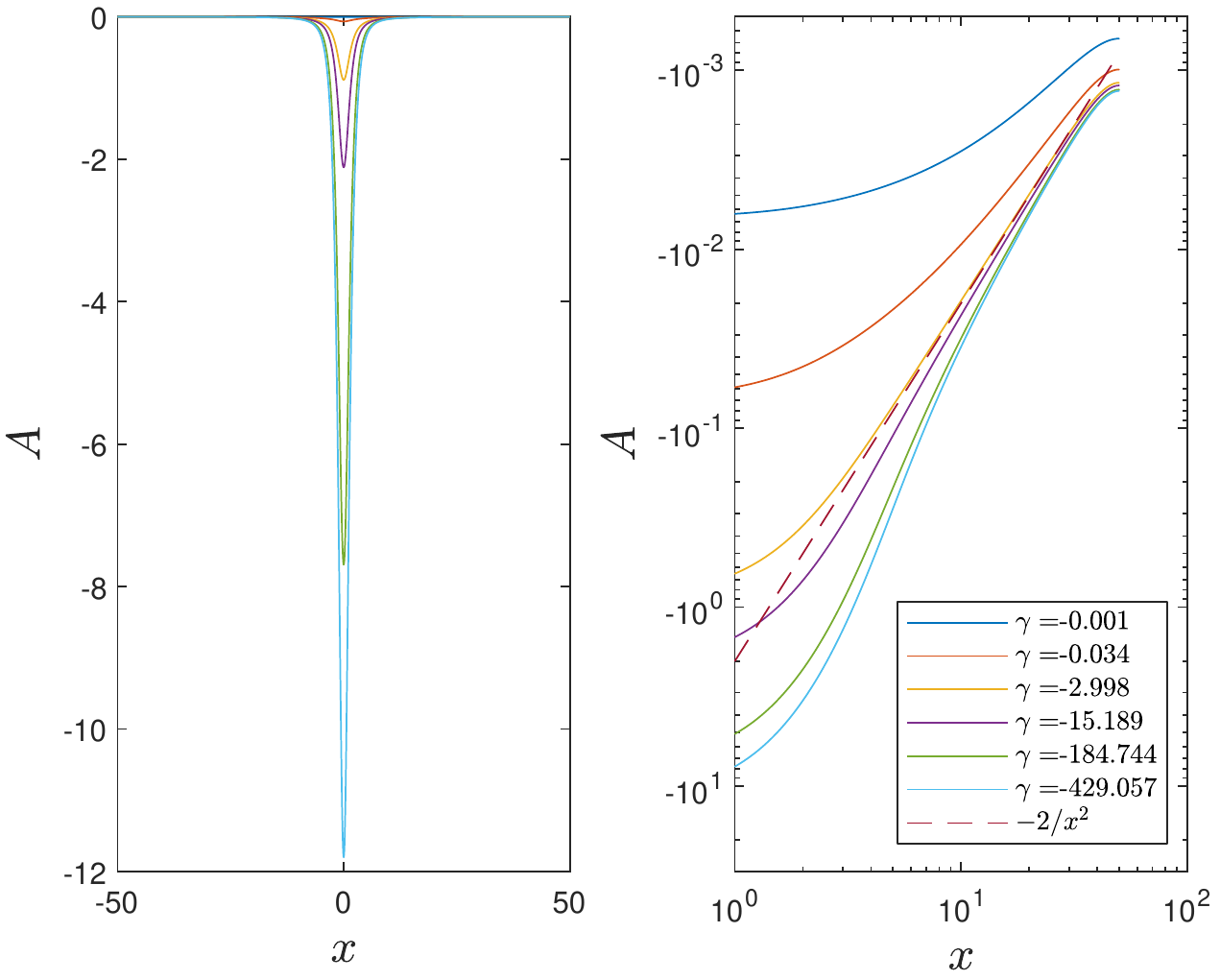}
		\caption{Evolution of Keeler type I solution to equation \eqref{sfkdv2} (left) and corresponding decay rate in log-log scale obtained at various $\gamma$ values (right). Behaviour approaches predicted rate in equation \eqref{algebraicdecay}. Results near $x=\pm50$ are unreliable due to Fourier methods enforcing periodic boundary conditions.}
		\label{decayrate}
	\end{figure}

	Figure \ref{decayrate} shows the far field decay rate of ADSW type I solution approaching the predicted decay rate of equation \eqref{predictdecay}. Similar behaviour is also observed for other ADSW solution types. 

	\subsection{Hydraulic parametric results}
	\label{subsection:Hydraulic parametric results}
	
\begin{figure}[H]
		\centering
		\begin{subfigure}[b]{0.49\textwidth}
			\includegraphics[width=\textwidth]{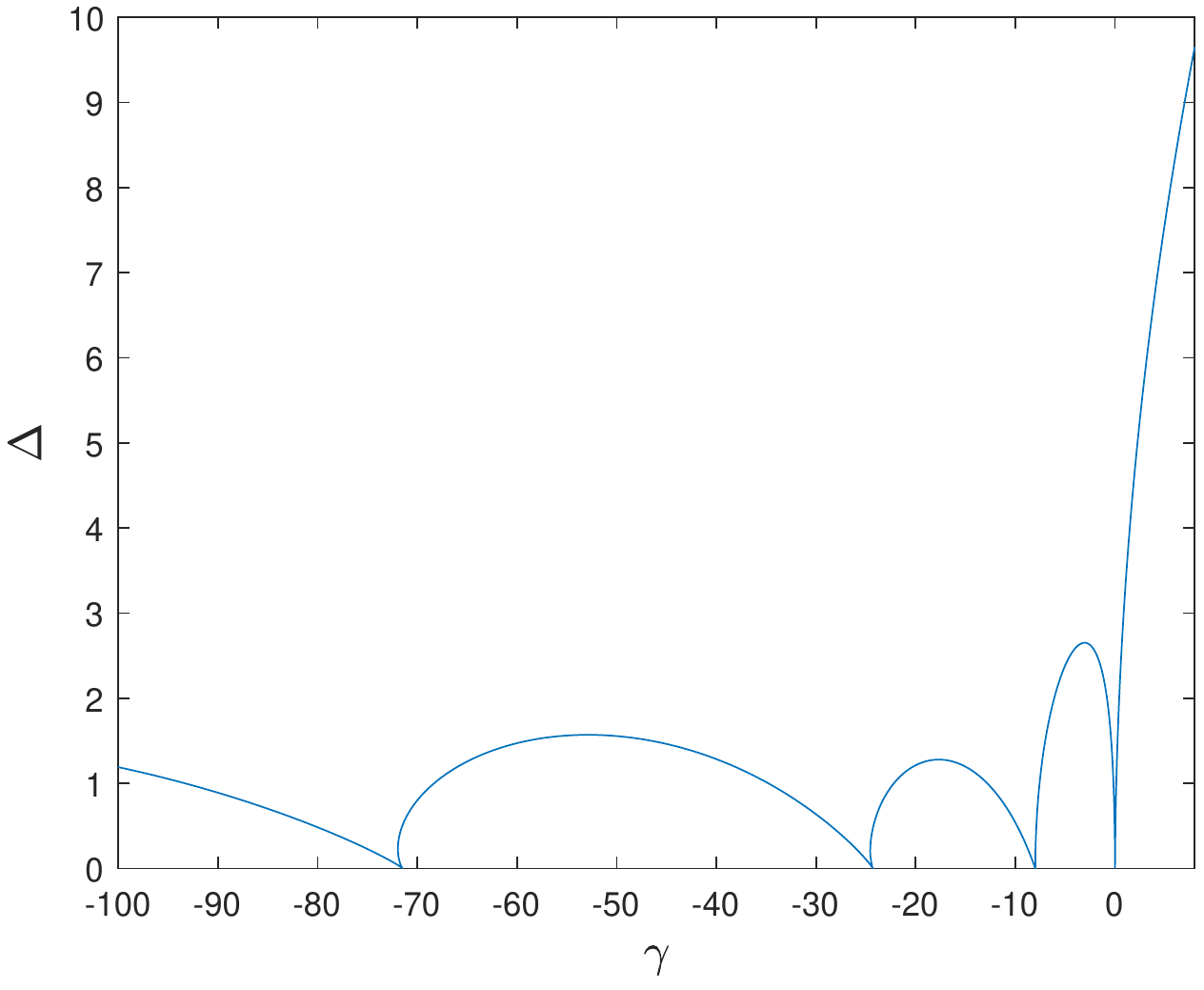}
		\end{subfigure}%
		\begin{subfigure}[b]{0.51\textwidth}
			\includegraphics[width=\textwidth]{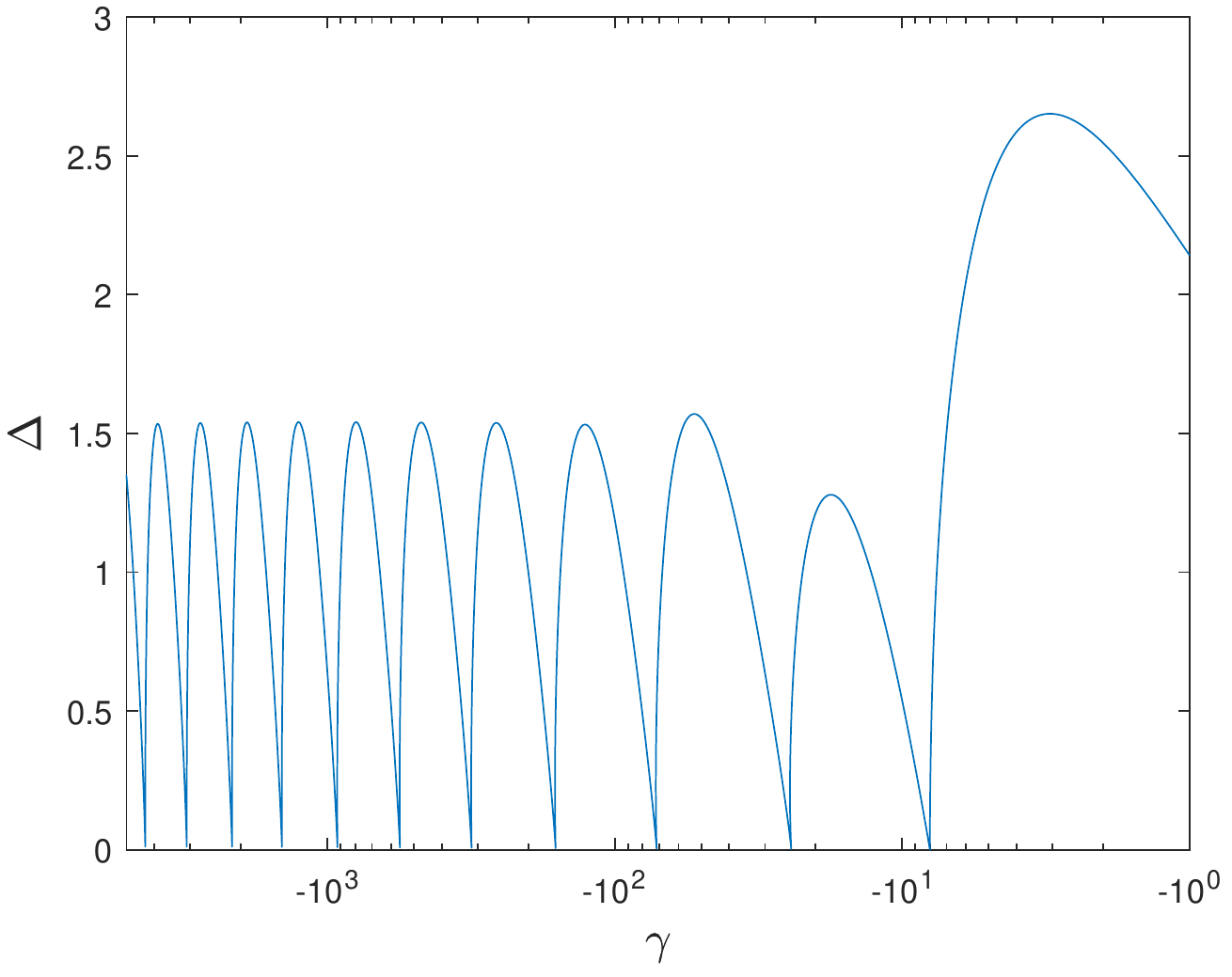}
		\end{subfigure}
		\caption{Parametric space for hydraulic fall solutions to equation \eqref{sfkdv} with $\gamma$ axis linear scaling (left) and log scaling (right). A small vertical turning point is present before every root (except at $\gamma=0,-8$)}
		\label{parametric}
	\end{figure}

	We now turn to extending the parametric space obtained by Ee and Clarke. The parametric relationship $\gamma(\Delta)$ for hydraulic fall solutions, generated using pseudo arc-length, is consistent with that obtained by Ee and Clarke in the region $\gamma\approx-24.5$ to $\gamma=10$. Our results also agree with the asymptotic analysis in the limit $\gamma\rightarrow0$, large positive $\gamma$ limit and the perturbation analysis around $\gamma=-8$. Furthermore, our numerical results are able to be significantly extended into the negative $\gamma$ region, as shown in Figure \ref{parametric}.\\

	Locations of each solitary wave solution coincide with the EDSW locations found using the shooting method. The EDSW solutions at $\Delta=0$ cannot be obtained using pseudo arc-length continuation, although solutions for very small $\Delta$ are available. This is likely due to the existence of a continuum of solutions on the line $\Delta=0$ shown by KBB resulting in Newton's method being unable to converge.\\
	
	An arch type behaviour is observed in the parametric space $\gamma(\Delta)$ between subsequent EDSW locations with a similar shape and height. There also appears to be a vertical turning point just before each negative eigenvalue (except at $\gamma=-8$) which had not been previously observed by Ee and Clarke. This is due to their minimisation algorithm being unable to handle vertical turning points. In contrast, pseudo arc-length has no such issues with traversing turning points. This is also the likely explanation for the small discrepancy at the second solitary wave location between Ee and Clarke, and Cullen.\\
	
	\section{Conclusion}
	\label{section:Conclusion}
	
	An improved numerical method has been developed to generate the extended hydraulic fall parameter space for solutions of the fKdV equation. This is accomplished using tabletop solutions obtained using twin topographies and a wave-drag constraint. Pseudo arc-length continuation is used with a SMG solver for each Newton iteration. Our method appears to be more numerically stable than the approach used by Ee and Clarke, and Cullen. We further show each EDSW in the parametric space for hydraulic fall solutions evolves into a separate type of ADSW continuum observed by KBB.\\
	
	Numerical evidence suggests that there may exist an infinite set of EDSW solutions to the fKdV equation along $\Delta=0$ as $\gamma\rightarrow-\infty$, corresponding to a nonlinear eigenvalue problem with eigenvalues $\gamma$ and EDSW eigenfunctions. The connection between EDSW and ADSW solutions further suggests the existence of infinitely many types of ADSW continuum solutions.\\
	
	Lastly, we show that hydraulic fall solutions periodically evolve into EDSW solutions. The hydraulic fall parametric relationship $\gamma(\Delta)$ follows a similar arch-type behaviour connecting subsequent EDSW.\\

	\bibliography{references}
	
\end{document}